\documentclass[aps,prl,twocolumn,nofootinbib,superscriptaddress,balancelastpage,letterpaper]{revtex4}

\usepackage[T1]{fontenc} % if needed
\usepackage[utf8]{inputenc}
\usepackage{subcaption}
\usepackage{graphicx}
\usepackage{graphics}
\usepackage[mathscr]{eucal}
\usepackage{amssymb} 
\usepackage{amsmath}
\usepackage{xspace}
\usepackage{listings}
\usepackage{ulem}
\usepackage{xcolor}
\usepackage{bm}         % for bold math characters in a math einvironment
\usepackage{array}
\usepackage{multirow}   % analogue to \multicolumn
\usepackage{booktabs}   % for fancy looking tables
\usepackage{mleftright}

\usepackage{bbold}
\usepackage{amsmath,amscd}
\usepackage{slashed}
\usepackage{placeins}
\usepackage{soul}
\usepackage{braket}
\usepackage{empheq}
\usepackage[most]{tcolorbox}
\usepackage{mathtools,slashed}
\usepackage{dsfont}
\usepackage[makeroom]{cancel}
\usepackage{multirow}
\usepackage{amssymb}

\usepackage{dcolumn}% Align table columns on decimal point
\usepackage{physics}
\usepackage{makecell} 

\usepackage{hyperref}
\hypersetup{colorlinks=true, linkcolor=blue, urlcolor=blue, citecolor=blue}

\usepackage[capitalise]{cleveref}
\crefname{section}{Sec.}{Secs.}
\crefname{table}{Tab.}{Tabs.}
\crefname{figure}{Fig.}{Figs.}
\crefname{equation}{Eq.}{Eqs.}
\crefname{appendix}{Appendix\ }{Appendix\ }

\newcommand{\U}[1]{\mathrm{U}(1)_{\mathrm{#1}}}			% Use this for U(1) groups
\renewcommand{\[}{\left[}

\newcommand\Tstrut{\rule{0pt}{2.6ex}}         % = `top' strut
\definecolor{ForestGreen}{rgb}{0.13, 0.55, 0.13}
\definecolor{Mulberry}{rgb}{0.77, 0.29, 0.55}
\definecolor{bostonuniversityred}{rgb}{0.8, 0.0, 0.0}
\definecolor{amber}{rgb}{1.0, 0.49, 0.0}

\newcommand\varpm{\mathbin{\vcenter{\hbox{%
  \oalign{\hfil$\scriptstyle\hspace{-0.1ex}+\hspace{-0.1ex}$\hfil\cr
          \noalign{\kern-.5ex}
          $\scriptscriptstyle({-})$\cr}%
}}}}

\begin{document}

\title{Supercooled phase transitions in conformal dark sectors explain NANOGrav data}

\author{João Gonçalves}
\email{jpedropino@ua.pt}
\affiliation{Departamento de F\'isica, Universidade de Aveiro and Centre for
Research and Development in Mathematics and Applications (CIDMA), Campus de Santiago, 
3810-183 Aveiro, Portugal}
\affiliation{Laboratório de Instrumentação e Física Experimental de Partículas (LIP), Universidade do Minho, 4710-057 Braga, Portugal}
\author{Danny~Marfatia}
\email{dmarf8@hawaii.edu}
\affiliation{Department of Physics and Astronomy, University of Hawaii at Manoa, Honolulu, HI 96822, USA}

\author{Ant{\'o}nio~P.~Morais}
\email{amorais@fisica.uminho.pt}
\affiliation{Departamento de Física, Escola de Ciências, Universidade do Minho, 4710-057 Braga, Portugal}
\affiliation{Laboratório de Instrumentação e Física Experimental de Partículas (LIP), Universidade do Minho, 4710-057 Braga, Portugal}
\affiliation{Departamento de F\'isica, Universidade de Aveiro and Centre for
Research and Development in Mathematics and Applications (CIDMA), Campus de Santiago, 
3810-183 Aveiro, Portugal}

\author{Roman~Pasechnik}
\email{Roman.Pasechnik@fysik.lu.se}
\affiliation{Department of Physics, Lund University, 221 00 Lund, Sweden}

\begin{abstract}
According to recent lore, it is difficult to explain the evidence for a stochastic gravitational wave background obtained by pulsar timing arrays with supercooled first-order phase transitions (FOPTs). We demonstrate that supercooled FOPTs in dark U(1)$'$ models with a conformal dark sector easily explain the nHz signal at NANOGrav. 
\end{abstract}

\maketitle
%%%%%%%%%%%%%%%%%%%%%%%%%%%%%%%%%%%%%%%%%%%%%%%%%%%%%%%%%%
\textit{\textbf{Introduction}.}
%%%%%%%%%%%%%%%%%%%%%%%%%%%%%%%%%%%%%%%%%%%%%%%%%%%%%%%%%%
The Standard Model (SM) is an unambiguously successful theory of particle physics. However, it fails to explain neutrino masses, dark matter, and the observed baryon asymmetry of the Universe. These shortcomings point to the need for new physics (NP) beyond the SM and motivate efforts to explore various theoretical avenues and synergies in the search for new phenomena.

Recently, pulsar timing arrays (PTAs) reported evidence of a stochastic gravitational wave background (SGWB) at nHz frequencies~\cite{NANOGrav:2023gor,EPTA:2023fyk,Reardon:2023gzh,Xu:2023wog}.  Statistical analyses have shown that NP scenarios are more compatible with the observed signal than the purely astrophysical explanation provided by supermassive black hole binaries (SMBHBs)~\cite{NANOGrav:2023hvm}. Supercooled MeV-scale first-order phase transitions (FOPTs) have been proposed as an explanation for the NANOGrav signal~\cite{NANOGrav:2023hvm,Ellis:2023oxs,Gouttenoire:2023bqy} and for signals at PTAs in general~\cite{Kobakhidze:2017mru}. However, in Ref.~\cite{Athron:2023mer} it is argued that such a scenario may not be physically viable because such FOPTs may fail to fully complete on the relevant cosmological time scales. It is further argued that even if this issue can be evaded, the Universe would be reheated to temperatures associated with the scale of NP driving the FOPT, effectively shifting the SGWB spectrum to frequencies higher than the NANOGrav signal.  However, these conclusions are obtained in the context of electroweak (EW) FOPTs in NP models with SM-like potentials that feature a cubic interaction term at tree level, thus raising the question about their validity for conformal NP models.

In this Letter, we demonstrate via a concrete counterexample how supercooled FOPTs in conformal dark sectors can circumvent the issues raised in Ref.~\cite{Athron:2023mer}. Specifically, we show that strongly supercooled FOPTs U(1)$'$ models with a conformal dark sector can explain the NANOGrav signal. 

%%%%%%%%%%%%%%%%%%%%%%%%%%%%%%%%%%%%%%%%%%%%%%%%%%%%%%%%%%
\textit{\textbf{Model}.} 
%%%%%%%%%%%%%%%%%%%%%%%%%%%%%%%%%%%%%%%%%%%%%%%%%%%%%%%%%%
We explore a dark conformal extension of the SM 
augmented by a gauged dark $\mathrm{U(1)}^\prime$ symmetry. The tree-level potential of the model is \begin{equation}\label{eq:tree_potential}
\begin{aligned}
V_{0} = &-\mu^2_h \, \mathcal{H}^\dagger \mathcal{H} + \lambda_h(\mathcal{H}^{\dagger}\mathcal{H})^2 + \lambda_{\sigma}(\sigma^{\dagger}\sigma)^2  + \\
&\lambda_{\sigma h}(\mathcal{H}^{\dagger}\mathcal{H}) (\sigma^{\dagger}\sigma)\,,
\end{aligned}
\end{equation}
which has scale invariance only in the dark sector. Upon spontaneous breaking of the EW and $\mathrm{U(1)}^\prime$ symmetries, at tree level and one loop, respectively, $\mathcal{H}$ and $\sigma$ acquire vacuum expectation values (vevs) $v\simeq 246~\mathrm{GeV}$ and $v_\sigma$. This yields two physical scalars -- the standard Higgs boson ($h_1$) and a dark scalar boson called a scalon ($h_2$) -- as well as the massive SM gauge bosons $W^\pm,\,\mathrm{Z^0}$ and a $\mathrm{Z^\prime}$ boson. 

Consider imposing a classical conformal symmetry in both the dark and visible sectors, {\it i.e.}, $\mu^2_h=0$, and requiring a large hierarchy between the vevs, $v_\sigma \gg v$, associated with a large hierarchy in the scalar mass spectrum, $M_{h_2} \gg M_{h_1}$~\cite{Goncalves:2024pr}. (Throughout, we fix the mass of the SM-like Higgs boson to its experimentally measured value $M_{h_1} = 125.11~\mathrm{GeV}$~\cite{ATLAS:2023oaq}.)
In this case, the scalon emerges from the heavy singlet $\sigma$ field. This is in variance with a multi-Higgs-doublet scenario which necessarily implies that the scalon is the lightest particle in the scalar spectrum~\cite{Gildener:1976ih}. Note that a large mass hierarchy between the scalon and the SM Higgs is protected against large radiative corrections by means of a small portal coupling $|\lambda_{\sigma h}| \sim (v/v_\sigma)^2$.  In this scenario, GW signals with peak frequencies above a mHz are obtained~\cite{Goncalves:2024pr}. 

On the other hand, an FOPT induced by the breaking of the gauged $\mathrm{U(1)}^\prime$ symmetry at an MeV scale could produce a GW signal in the nHz range relevant to PTA observations. However, in this case, with the opposite hierarchy $v \gg v_\sigma$, corresponding to $M_{h_1} \gg M_{h_2}$, the scalon emerges from $\mathcal{H}$. Now there are two possibilities: 1) If $\lambda_h < y_t$, where $y_t$ is the top Yukawa coupling, the top quark loop is the dominant contribution to the Higgs mass, which comes with a negative sign and alters its vacuum stability, such that no valid solution exists. 2)~If $\lambda_h > y_t$, the scalar loop is the dominant contribution and valid solutions are possible. However, as shown in Ref.~\cite{Elias:2003zm}, the Higgs mass predicted by minimizing the effective potential exceeds 200~GeV, which is inconsistent with experiment.

We avoid this problem by resorting to a novel approach where only the dark sector is conformal {\it i.e.}, the $\sigma^{\dagger}\sigma$ term is absent, but not the visible one ($\mu^2_h\not=0$). This is necessary to ensure proper EW symmetry breaking with $M_{h_1} \gg M_{h_2}$. The Coleman-Weinberg (CW) mechanism~\cite{Coleman:1973jx} of radiative $\mathrm{U(1)}^\prime$ symmetry breaking is analyzed at one-loop order. Following Ref.~\cite{Goncalves:2024pr}, the potential in Eq.~\eqref{eq:tree_potential} is supplemented by the one-loop CW potential \cite{Coleman:1973jx}, {\it i.e.}, $V = V_0 + V_{\mathrm{CW}}$, enabling us to also compute the mass spectrum at one loop. This procedure fixes the values of $\lambda_h$, $\lambda_\sigma$, $\mu^2_h$ and the singlet vev $v_\sigma$, leaving the scalon mass $M_{h_2}$ and the scalar mixing $\lambda_{\sigma h}$ as the physical input parameters of the scalar sector. However, the role of $\lambda_{\sigma h}$ is rendered insignificant by the quadratic Higgs term. In the gauge sector, the additional Abelian gauge field mixes kinetically with the hypercharge $\mathrm{U(1)}_{\rm Y}$ field. This can be parameterized by a gauge coupling $g_{12}$ in addition to the $\mathrm{U(1)}^\prime$ gauge coupling $g_L$. 
We fix the value of $g_{12}$ as described below.
Hence, the only additional free parameter arising
from the gauge sector is $g_L$.
In the renormalization group (RG) evolution, we require that $\lambda_h$ remains positive and that all couplings are perturbative up to a renormalization scale $\mu=10^{8}~\mathrm{GeV}$, which sets the scale to which our model is valid.

Equipped with this theoretical framework, we turn to a discussion of its cosmological implications. To satisfy the constraints on the extra effective number of neutrino species, $\Delta N_\mathrm{eff}$, we assume that the dark sector thermalizes with the visible sector and is non-relativistic after the phase transition. 
For MeV mass scales, we ensure that $\mathrm{Z^\prime}$ thermalizes by fixing $g_{12} = 2 \times 10^{-10}$ at the EW scale~\cite{Redondo:2008ec}. On the other hand, $h_2$ can thermalize through the   $h_2\mathrm{Z^\prime}\mathrm{Z^\prime}$ vertex which is proportional to $g_L \sim \mathcal{O}(0.1-1)$. Furthermore, the total integrated SGWB energy density $h^2 \Omega_\mathrm{GW}$ must not exceed the amount of dark radiation allowed during the epoch of Big Bang Nucleosynthesis. (The Hubble constant $H_0 = 100h~\mathrm{km/s/Mpc}$.) Then, $h^2 \Omega_\mathrm{GW} < 5.6 \times 10^{-6} \Delta N_\mathrm{eff}$ \cite{Pagano:2015hma}, which yields $h^2 \Omega_\mathrm{GW} < 2.8 \times 10^{-6}$ for $\Delta N_\mathrm{eff} < 0.5$.

%%%%%%%%%%%%%%%%%%%%%%%%%%%%%%%%%%%%%%%%%%%%%%%%%%%%%%%%%%
\textit{\textbf{RG-improved thermal potential}.}
%%%%%%%%%%%%%%%%%%%%%%%%%%%%%%%%%%%%%%%%%%%%%%%%%%%%%%%%%%
RG evolution significantly impacts SGWB spectra~\cite{Gould:2021oba,Croon:2020cgk}. In fact, small variations in the RG scale can cause substantial changes in the SGWB amplitude. To address this issue, we consider an RG-improved potential~\cite{Kierkla:2022odc,Chataignier:2018kay,Chataignier:2018aud,Ellis:2020nnr}, wherein the couplings and the classical field configuration $\phi_\sigma$ in the potential satisfy the RG equations and are replaced by~\cite{Goncalves:2024pr} 
\begin{equation}\label{eq:RG_transformations}
\begin{aligned}
&\lambda_i \rightarrow \lambda_i(t)\,, \ \ 
&\phi_\sigma^2 \rightarrow \frac{\phi_\sigma^2}{2} \exp{\int_0^t  3g_L^2(t^\prime)dt' }\,,
\end{aligned}
\end{equation}
where $\lambda_i = \mu_h^2\,,\lambda_\sigma\,,\lambda_h\,,\lambda_{\sigma h}, y_t, g_L, g_{12}$. Here, $t \equiv \mathrm{ln}(\mu/\mu_{\mathrm{ref}})$, with the reference scale set to $\mu_{\mathrm{ref}} \equiv M_{\mathrm{Z^0}} = 91~\mathrm{GeV}$. Different choices of $\mu_{\mathrm{ref}}$ have a minimal impact on the shape of the effective potential. The couplings $\lambda_i(t)$ evolve according to the $\beta$ functions provided in the Appendix.

Since we focus on phase transition temperatures much lower than the EW scale, we do not consider contributions from $W^\pm$ and $\mathrm{Z^0}$ in the RG evolution. Also, we only consider the $\sigma$ direction to be relevant for the FOPT as it decouples from the Higgs direction due to the strong scale hierarchy $M_{h_1} \gg M_{h_2}$. The field-dependent RG scale is chosen to be $\mu = \mathrm{max}[M_{\mathrm{Z^\prime}}(\phi_\sigma), \pi T]$, where $M_{\mathrm{Z^\prime}}(\phi_\sigma)$ is the field-dependent $\mathrm{Z^\prime}$ mass,
\begin{equation}\label{eq:massZprime}
    M_{\mathrm{Z}^{\prime}}^2=\frac{1}{4}g_{12}^2 v^2+g_L^2 \phi_\sigma^2\,.
\end{equation}
Thermal corrections to the potential are introduced through the one-loop thermal integrals $V_T$~\cite{Quiros:1999jp}, as well as through the Daisy resummation term, $V_{\mathrm{Daisy}}$ \cite{Arnold:1992rz}. The full effective potential is then given by $V_{\mathrm{eff}} = V_0(\phi_\sigma) + V_{\mathrm{CW}}(\phi_\sigma) + V_T(\phi_\sigma,T) + V_{\mathrm{Daisy}}(\phi_\sigma,T)$, where all the couplings and fields are redefined in accordance with Eq.~\eqref{eq:RG_transformations} for the zero and finite temperature parts. Explicit expressions for $V_T$ and $V_{\mathrm{Daisy}}$ can be found in Ref.~\cite{Goncalves:2024pr}. 

%%%%%%%%%%%%%%%%%%%%%%%%%%%%%%%%%%%%%%%%%%%%%%%%%%%%%%%%%%
\textit{ \textbf{Gravitational waves}.}
%%%%%%%%%%%%%%%%%%%%%%%%%%%%%%%%%%%%%%%%%%%%%%%%%%%%%%%%%%
 The thermodynamic parameters needed to determine the SGWB spectrum are computed from the tunneling (bounce) action between the true and false vacuum using \texttt{CosmoTransitions} \cite{Wainwright:2011kj}. The SGWB spectrum can be characterized by: 1) the percolation temperature $T_p$, corresponding to the temperature when 34\% of the Universe is filled by the true vacuum; 2) the reheating temperature $T_\mathrm{RH}$ at the end of the phase transition; 3) the strength of the FOPT $\alpha$, defined as the ratio of the latent heat released to the total radiation energy density; 4) the inverse time duration of the FOPT normalized to the Hubble rate $\beta/H$, and 5) the bubble wall velocity $v_w=1$ for supercooled phase transitions. These five parameters can then be used in spectral templates to determine the frequency and amplitude of the SGWB signal. We utilized the latest templates provided by the LISA Cosmology working group~\cite{Caprini:2024hue} for bubble collisions and sound waves in the plasma as the main sources of GWs.

%%%%%%%%%%%%%%%%%%%%%%%%%%%%%%%%%%%%%%%%%%%%%%%%%%%%%%%%%%
\textit{ \textbf{FOPTs in conformal dark sectors}.}
%%%%%%%%%%%%%%%%%%%%%%%%%%%%%%%%%%%%%%%%%%%%%%%%%%%%%%%%%%
\begin{figure}[t!]
	\centering
    \captionsetup{justification=raggedright, singlelinecheck=false}
\subfloat{\includegraphics[width=0.5\textwidth]{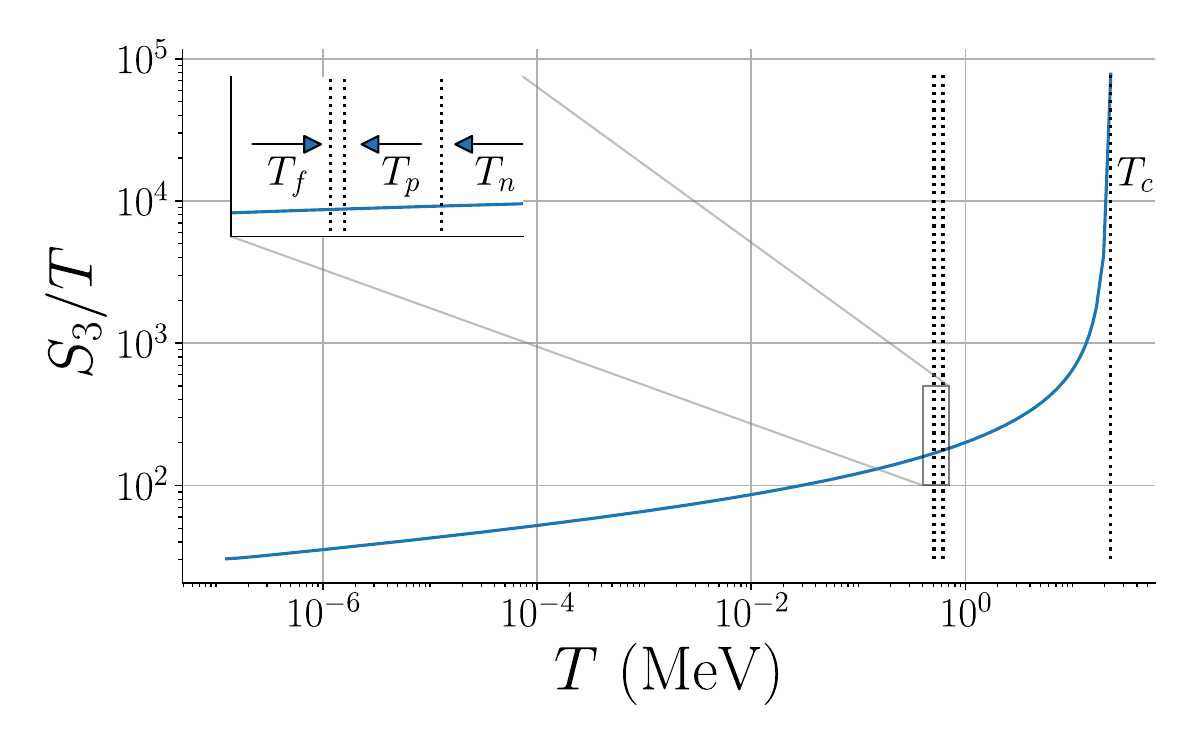}}
    \caption{Ratio of the Euclidean action to temperature $S_3/T$ as a function of $T$ for the best-fit point in Table~\ref{tab:benchmarks}. The vertical dotted lines indicate the critical temperature $T_c$, nucleation temperature $T_n$, percolation temperature $T_p$, and completion temperature $T_f$. }
	\label{fig:action_vs_temp}
\end{figure}
\begin{figure}[t!]
	\centering
    \captionsetup{justification=raggedright, singlelinecheck=false}
	\subfloat{\includegraphics[width=0.5\textwidth]{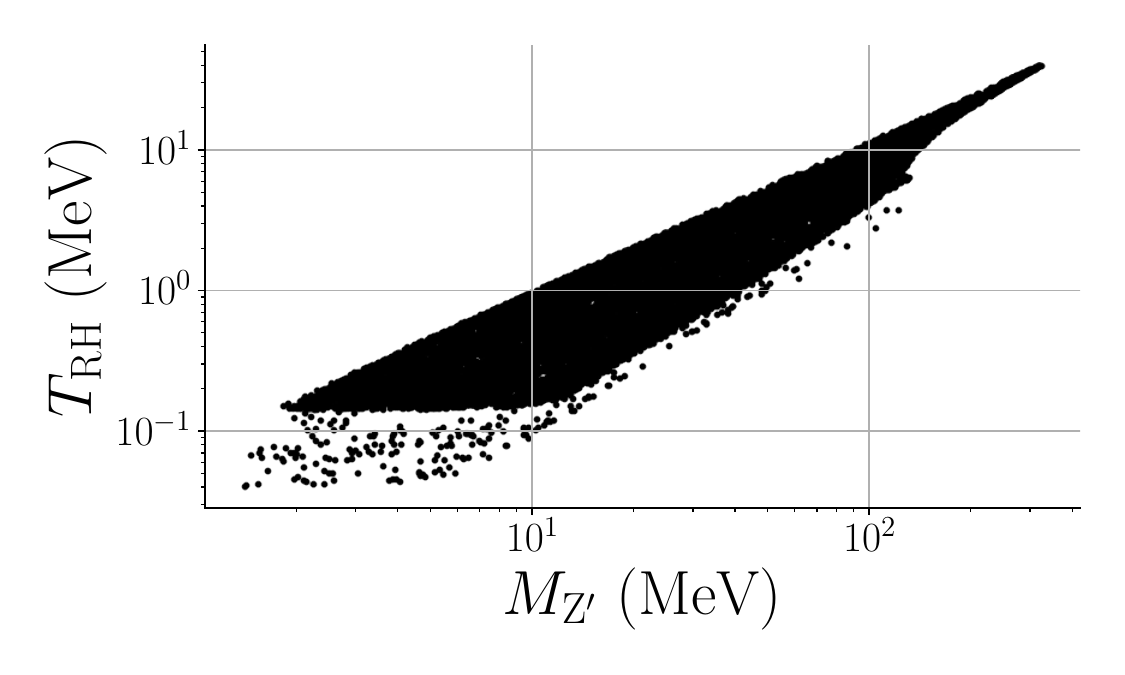}}
    \caption{Scatter plot of the reheating temperature $T_\mathrm{RH}$ as a function of the $\mathrm{Z^\prime}$ mass.}
	\label{fig:TRH_vs_mZp}
\end{figure}
FOPTs in conformal models are typically supercooled as the potential barrier between the true and false vacuum persists as $T \rightarrow 0$ allowing for percolation to occur at arbitrarily low temperatures.

Our scenario overcomes the issues highlighted in Ref.~\cite{Athron:2023mer}. First, in conformal models percolation is always possible. We have explicitly verified that the volume of the false vacuum is decreasing at the percolation temperature $T_p$ by requiring
\begin{equation}\label{eq:perc_condition}
H(T)\left(3 + T\frac{dI}{dT}\right)\Biggr|_{\substack{T = T_p}} < 0\,,
\end{equation}
where $H(T)$ is the Hubble parameter and $I(T)$ is the true vacuum volume per unit comoving volume. As discussed in Ref.~\cite{Goncalves:2024pr}, this condition may not necessarily be satisfied at $T_p $, but at some lower temperature. However, here we only consider FOPTs for which the percolation condition is fulfilled at $T = T_p$. Second, in non-conformal models, such as the type studied in Ref.~\cite{Athron:2023mer}, the three-dimensional Euclidean action as a function of temperature exhibits a $U$-shaped behavior (see {\it e.g.}, Fig.~3 of Ref.~\cite{Levi:2022bzt}), which bounds the percolation temperature from below. In contrast, conformal models have an action that asymptotically approaches zero as $T \rightarrow 0$; see Fig.~\ref{fig:action_vs_temp}. Third, we have checked that the potential remains bounded from below at the FOPT completion temperature $T_f$, at which the true vacuum occupies 99\% of the Universe. Fourth, the conformal nature of the potential prevents $T_\mathrm{RH} \approx M$, where $M$ is the mass scale of dark sector. Since the potential barrier persists as $T \rightarrow 0$, the FOPT can be delayed to temperatures well below $T_c$, which defines the energy scale driving the transition. Consequently, as long as extreme supercooling is avoided ($\alpha \lesssim 10^8$), $T_\mathrm{RH}$ deviates from $T_p$ by at most a factor of $\mathcal{O}(1-100)$, ensuring that $T_\mathrm{RH}<M$; see Fig.~\ref{fig:TRH_vs_mZp}. 
(It was also noted in Ref.~\cite{Ellis:2020nnr} that $T_\mathrm{RH} \sim 0.1 M_\mathrm{Z^\prime}$ for a conformal $\U{B-L}$ model.)
In this regime, Daisy corrections are negligible. 

%%%%%%%%%%%%%%%%%%%%%%%%%%%%%%%%%%%%%%%%%%%%%%%%%%%%%%%%%
\textit{\textbf{Numerical results}.}
%%%%%%%%%%%%%%%%%%%%%%%%%%%%%%%%%%%%%%%%%%%%%%%%%%%%%%%%%
\begin{figure}[t]
	\centering
    \captionsetup{justification=raggedright, singlelinecheck=false}
	\subfloat{\includegraphics[width=0.48\textwidth]{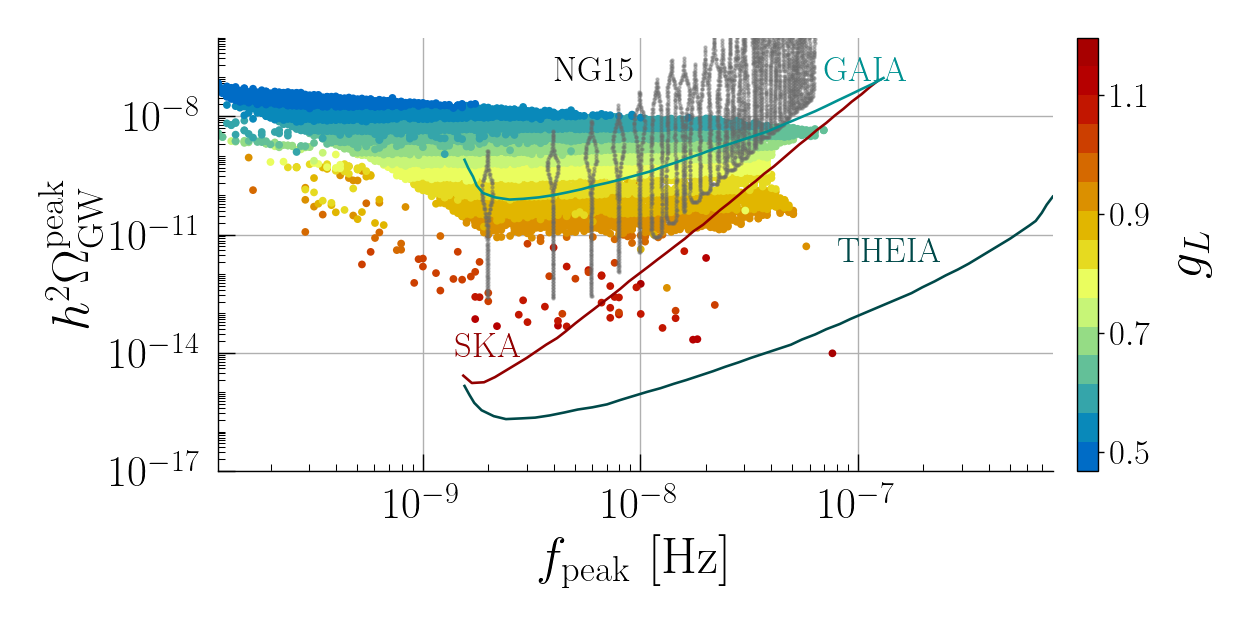}} \\
    \subfloat{\includegraphics[width=0.48\textwidth]{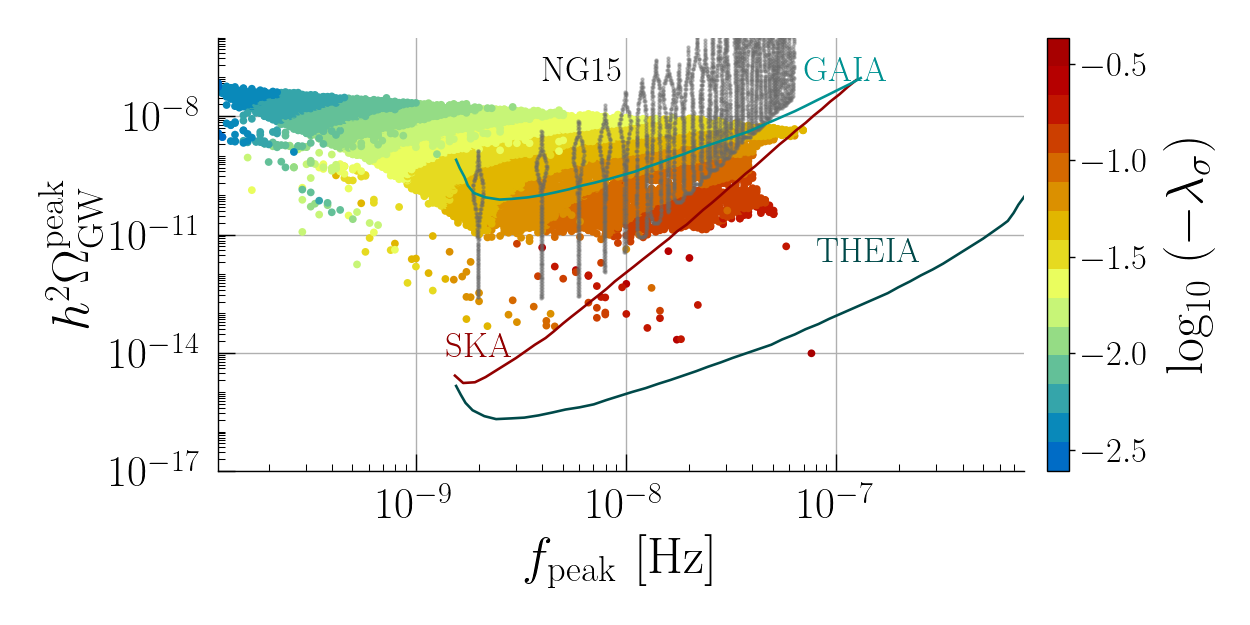}} \\
	\subfloat{\includegraphics[width=0.48\textwidth]{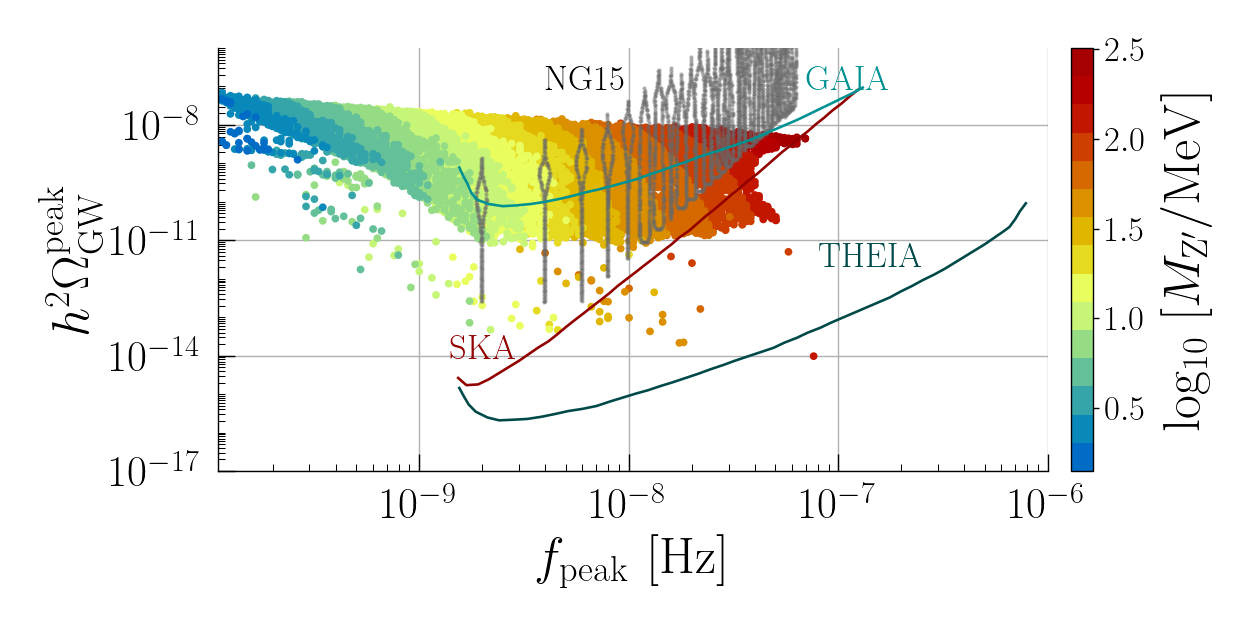}}
    \caption{Scatter plots of the SGWB peak amplitude $h^2 \Omega_\mathrm{GW}^\mathrm{peak}$ versus peak frequency $f_\mathrm{peak}$. The color scales indicate the gauge coupling $g_L$, the scalar self-interaction $\lambda_\sigma$ and the $\mathrm{Z^\prime}$ mass. For the NANOGrav signal (labeled "NG15") and sensitivity curves, the axis labels do not correspond to peak values. }
	\label{fig:proj_ompeak_fpeak}
\end{figure}
We perform a scan over the model parameters at $\mu=M_\mathrm{Z^0}$ with a focus on the parameter space capable of explaining the NANOGrav data. Since we focus on MeV scale FOPTs, we sample $M_{h_2}$ logarithmically in the range $M_{h_2} = [0.1, 100]~\mathrm{MeV}$. We require $|\lambda_{\sigma h}| < 10^{-10}$ so that $v_\sigma$ is small enough to yield a signal at PTAs. For these values of $\lambda_{\sigma h}$, the impact on the FOPT is negligible.
The gauge coupling is sampled linearly in the range \mbox{$g_L = [0.25,1.5]$}. In what follows, all parameters are evaluated at $\mu = 0.1~\mathrm{MeV}$ after RG evolution.
 Figure~\ref{fig:proj_ompeak_fpeak} displays scatter plots of the SGWB peak amplitude versus peak frequency for $g_L$ (top panel), $\lambda_\sigma$ (middle panel) and $M_\mathrm{Z^\prime}$ (bottom panel). Also shown is the NANOGrav signal~\cite{NANOGrav:2023gor} and the sensitivity of the experiments, GAIA, SKA, and THEIA~\cite{Garcia-Bellido:2021zgu,Weltman:2018zrl}. Our analysis indicates that the peak frequency is primarily determined by $M_\mathrm{Z^\prime}$, with smaller masses corresponding to lower frequencies. $M_{h_2}$ exhibits a similar correlation and is not shown. The couplings $-\lambda_\sigma$ and $g_L$ influence the peak amplitude, with smaller values resulting in higher amplitudes. This is because they affect the location of the minimum of the effective potential and its depth~\cite{Goncalves:2024pr}. We also require $T_\mathrm{RH}<M$ to prevent extreme supercooling, as greater supercooling results in higher $T_\mathrm{RH}$, which shifts the GW spectrum to higher frequencies. The isolated points with $h^2 \Omega_\mathrm{GW}^\mathrm{peak} < 10^{-11}$ correspond to non-supercooled FOPTs with $\alpha < 0.1$ and $g_L  > 1.1$.

\begin{figure}[t!]
	\centering
    \captionsetup{justification=raggedright, singlelinecheck=false}
	\subfloat{\includegraphics[width=0.48\textwidth]{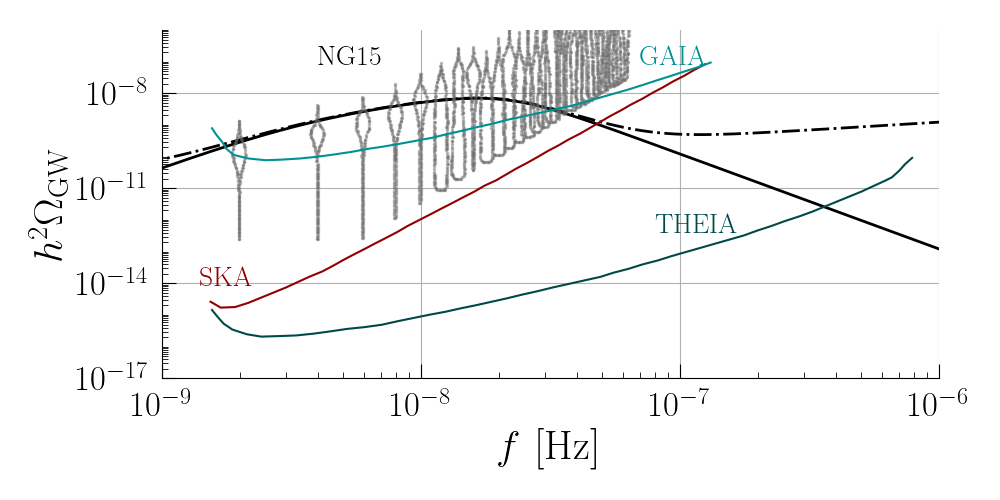}}
    \caption{SGWB spectra for the best-fit point in Table~\ref{tab:benchmarks}. 
    The solid black curve is the signal from the FOPT only, and the dot-dashed curve also includes the SMBHB signal.}
	\label{fig:Spectrum_ng5}
\end{figure}

The SGWB spectrum for the best-fit point, including the expected signal from SMBHBs, is shown in Fig.~\ref{fig:Spectrum_ng5}. The corresponding parameters are listed in Table~\ref{tab:benchmarks}. The SMBHB signal is modeled as a power law 
with spectral index $\gamma_\mathrm{BHB}$ and amplitude $A_\mathrm{BHB}$~\cite{Mitridate:2023oar}:
\begin{equation}
h^2 \Omega_{\rm {GW}}(f)={\frac{2\pi^2 A_{\rm {BHB}}^2}{3H_0^2}}\bigg(\frac{f}{\rm{year}^{-1}}\bigg)^{5-\gamma_{\rm BHB}} {\rm {year}}^{-2}\,.
\end{equation}
Clearly, the SMBHB contribution to the signal is negligible.

\begin{table}[t]
	\centering
    \captionsetup{justification=raggedright, singlelinecheck=false}
	\begin{tabular}{|c|c|c|c|c|c|c|c|}
		 $g_L$ & $M_{h_2}$ & $M_{\mathrm{Z^\prime}}$ & $T_\mathrm{RH}$ & $\alpha/10^5$ & $\beta/H$ & $\gamma_\mathrm{BHB}$ & $A_\mathrm{BHB}$ \\[0.5em] \hline\Tstrut
	   $0.59$ & $12.4$ & $107.3$ & $11.7$ & $2.60$ & $39.5$ & $4.50$ & $10^{-15.4}$
	\end{tabular} %174
	\caption{Parameters for the best fit point (including the contribution from SMBHBs) shown in Fig.~\ref{fig:Spectrum_ng5}. $M_{h_2}$, $M_{\mathrm{Z^\prime}}$ and $T_\mathrm{RH}$ are in units of MeV, and evaluated at $\mu=0.1$~MeV.} \label{tab:benchmarks}
\end{table}

\begin{figure}[t]
	\centering
    \captionsetup{justification=raggedright, singlelinecheck=false}
    \subfloat{\includegraphics[width=0.24\textwidth]{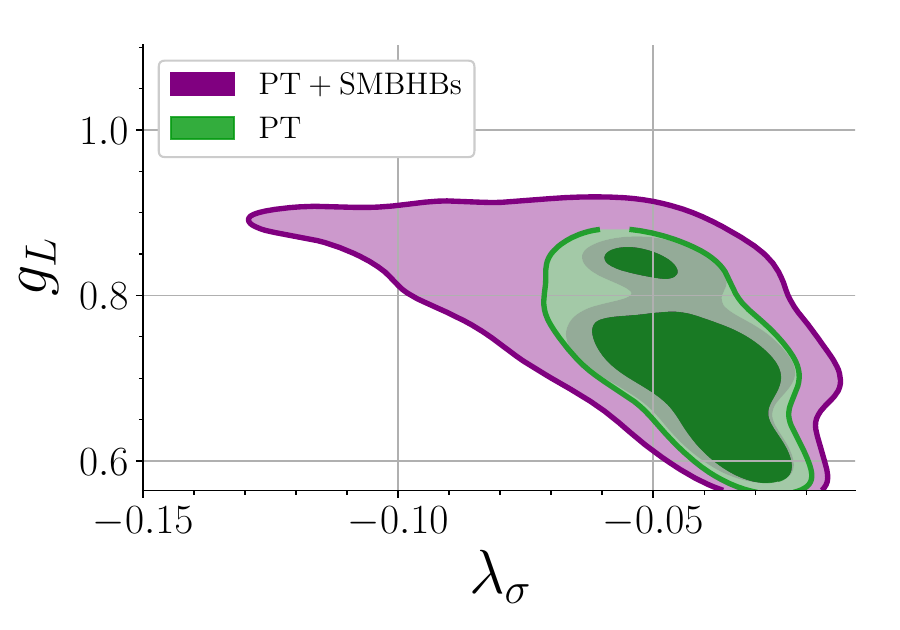}} 
    %\subfloat{\includegraphics[width=0.24\textwidth]{ls_vs_lsh_new.pdf}} \\
	\subfloat{\includegraphics[width=0.24\textwidth]{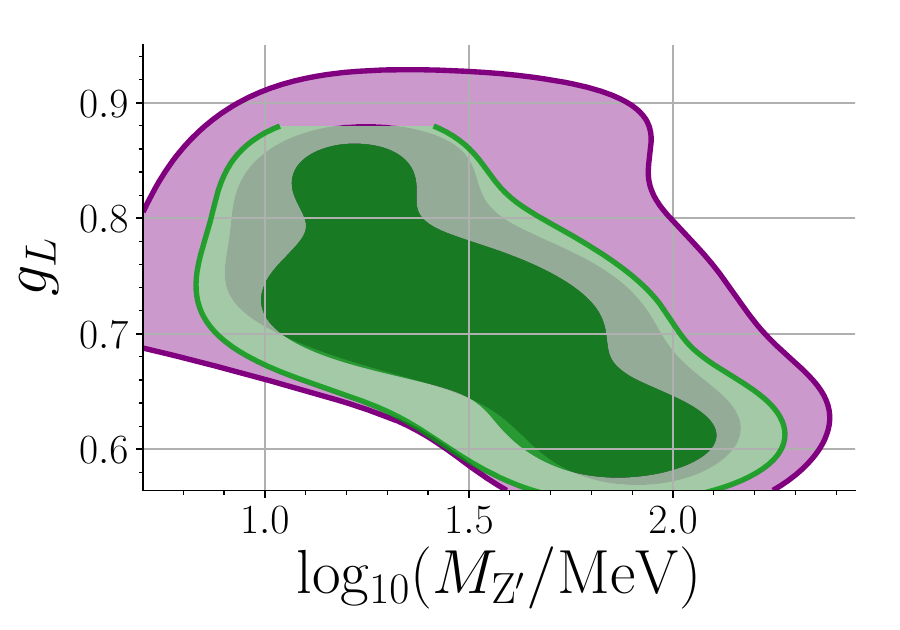}} \\
    \subfloat{\includegraphics[width=0.24\textwidth]{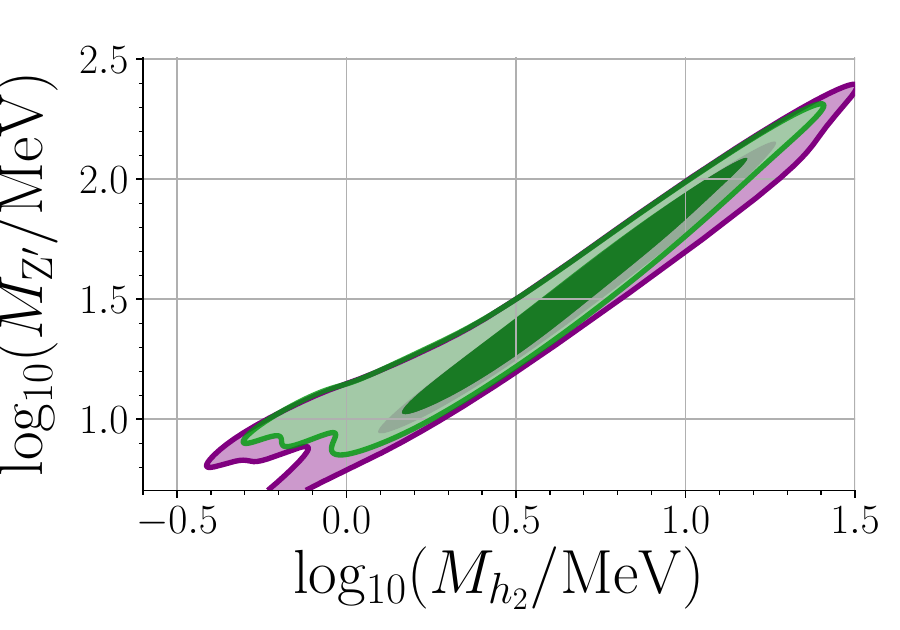}} 
    \caption{68\% CL (darker) and 95\% CL (lighter) allowed regions obtained from the NANOGrav 15-year dataset. The green contours assume the FOPT alone explains the signal, while the purple contours include the SMBHB signal. }	\label{fig:posterior_indiv_proj}
\end{figure}

To determine the 68\% and 95\% confidence level (CL)  parameter space regions allowed by the NANOGrav 15-year dataset, we employ the \texttt{PTArcade} package in its default \texttt{ceffyl} configuration~\cite{Mitridate:2023oar, Lamb:2023jls}. Figure~\ref{fig:posterior_indiv_proj} shows allowed regions assuming either that the model fully explains the signal (labeled "PT") or that the contribution from SMBHBs is included (labeled "PT+SMBHBs"). For the SMBHB signal, we adopt the default settings of \texttt{PTArcade}, choosing a uniform prior $\gamma_\mathrm{BHB} = [0,7]$ and a log-uniform prior $\mathrm{log_{10}} A_\mathrm{BHB} = [-18,-14]$. NANOGrav data favor $\mathrm{Z^\prime}$ and $h_2$ masses in the 10--100~MeV range at 68\% CL, with larger masses disfavored due to the resultant higher peak frequencies. We observe a strong correlation between $\lambda_\sigma$ and $g_L$. Smaller magnitudes are favored because they yield larger peak amplitudes. Without the SMBHB contribution, we find $g_L \in [0.57, 0.86]$ and $\lambda_\sigma \in [-0.062,-0.028]$ at 68\% CL.  As expected, including the SMBHB signal widens the allowed parameter regions, although in most cases the widening is minimal because the corresponding parameters show weak correlations with either the amplitude, frequency, or both.
%

%%%%%%%%%%%%%%%%%%%%%%%%%%%%%%%%%%%%%%%%%%%%%%%%%%%%%%%%%%
\textit{\textbf{Summary}.} 
%%%%%%%%%%%%%%%%%%%%%%%%%%%%%%%%%%%%%%%%%%%%%%%%%%%%%%%%%%
We showed that a dark conformal extension of the SM augmented by a dark $\mathrm{U(1)}^\prime$ symmetry can produce a supercooled FOPT that explains the GW signal observed by NANOGrav. The model is successful provided the FOPT is not extremely supercooled ($\alpha \lesssim 10^8$).
%This provides a counterexample to the expectation that supercooled FOPTs are unable to explain NANOGrav data. 
The pitfalls noted in Ref.~\cite{Athron:2023mer} are overcome by ensuring that the phase transition completes and that the dark sector remains non-relativistic after reheating. We also obtained model parameter values favored by the NANOGrav 15-year data. Specifically, dark sector masses of $\mathcal{O}(1-100)~\mathrm{MeV}$, couplings $g_L$ of $\mathcal{O}(0.1)$ and $\lambda_\sigma$ of $\mathcal{O}(0.01)$ are favored. Interestingly, NANOGrav data require $|\lambda_{\sigma h}|<10^{-10}$.

\vspace{0.1in}
%\newpage
\begin{acknowledgments}
\textit{{Acknowledgments}.}
J.G. and A.P.M. were supported supported by the Center for Research and Development in Mathematics and Applications (CIDMA) under the Portuguese Foundation for Science and Technology (FCT - Funda\c{c}\~{a}o para a Ci\^{e}ncia e a Tecnologia) Multi-Annual Financing Program for R\&D Units.
J.G. is also directly funded by FCT through the doctoral program grant with the reference 2021.04527.BD (\url{https://doi.org/10.54499/2021.04527.BD}).
D.M. is supported in part by the U.S. Department of Energy under Grant No.DE-SC0010504.
R.P.~and J.G.~are supported in part by the Swedish Research Council grant, contract number 2016-05996. R.P.~also acknowledges support by the COST Action CA22130 (COMETA).

\end{acknowledgments}

\appendix
\section{Appendix: Renormalization group equations}\label{app:rges}

The beta functions for the dark $\mathrm{U(1)^\prime}$ model,
\begin{equation*}
\beta\left(X\right) \equiv \mu \frac{d X}{d \mu}\equiv \frac{1}{16 \pi^2}\beta^{(1)}(X)\,,
\end{equation*}
 are given by
\begin{align}\label{eq:beta_functions}
    \begin{split}
    &\beta^{(1)}(g_L) = \frac{41}{10}g_{12}^2 g_L + \frac{g_L^3}{3}\,,
    \end{split} \\
    \begin{split}
    & \beta^{(1)}(g_1) = \frac{41}{10} g_1 (g_1^2 + g^2_{12})\,, 
    \end{split} \\
    \begin{split}
    & \beta^{(1)}(g_2) = -\frac{19}{6} g^2_2\,, 
    \end{split} \\
    \begin{split}
    & \beta^{(1)}(g_3) = -7 g^2_3\,, 
    \end{split} \\
    \begin{split}
    & \beta^{(1)}(g_{12}) = \frac{41}{10}g_{12} (g_1^2 + g_{12}^2) + \frac{1}{2} g_{12} g_L^2\,, 
    \end{split} \\
    \begin{split}
    & \beta^{(1)}(y_t) = \frac{3}{2}y_t^3 + y_t \Big[-\frac{17}{20} g_1^2 -\frac{17}{20} g_{12}^2 -\frac{9}{4} g_2^2 - \\
    &\hspace{5em} 8 g_3^2 + 3 y_t^2\Big]\,, 
    \end{split} \\
    \begin{split}
    & \beta^{(1)}(\lambda_h) = \frac{27}{200}g_1^4 + \frac{9}{20}g_1^2g_2^2 + \frac{9}{8}g_2^4 - \frac{9}{5}g_1^2\lambda_h - \\
    &\hspace{5em} \frac{9}{5} g_{12}^2\lambda_h - 9 g_2^2 \lambda_h + 24 \lambda_h^2 + \lambda_{\sigma h}^2 + \\
    &\hspace{5em} 12 \lambda_h y_t^2 - 6 y_t^4\,, 
    \end{split}  \\
    \begin{split}
    & \beta^{(1)}(\lambda_\sigma) = 2\Big[3g_L^4 - 6g_L^2\lambda_\sigma + 10 \lambda_\sigma^2 + \lambda_{\sigma h}^2\Big]\,, 
    \end{split} \\
    \begin{split}
    & \beta^{(1)}(\lambda_{\sigma h}) = \frac{\lambda_{\sigma h}}{10}\Big[-9g_1^2 - 9g_{12}^2 - 45 g_2^2 - 60g_L^2 + \\
    &\hspace{5em} 120\lambda_h + 80\lambda_\sigma + 40 \lambda_{\sigma h} + 60y_t^2\Big]\,, 
    \end{split} \\
    \begin{split}
    & \beta^{(1)}(\mu_h^2) = \frac{\mu_h^2}{10} \Big[-9g_1^2 - 9 g_{12}^2 - 45 g_2^2 + 120 \lambda_h + \\
    &\hspace{5em} 60y_t^2\Big]\,.
    \end{split}    
\end{align}

\bibliography{Refs.bib}

\end{document}